\documentclass[%
superscriptaddress,
preprint,
 amsmath,amssymb,
 aps,
]{revtex4-2}

\usepackage{graphicx}
\usepackage{dcolumn}
\usepackage{bm}

\begin{document}


\title{Predicting regional COVID-19 hospital admissions in Sweden using mobility data}

\author{Philip Gerlee}
 \email{gerlee@chalmers.se} 
\affiliation{Mathematical Sciences, Chalmers University of Technology and University of Gothenburg, Sweden}

\author{Julia Karlsson}
\affiliation{
 Sahlgrenska University Hospital, Gothenburg, Sweden
}%
\author{Ingrid Fritzell}
\author{Thomas Brezicka}
\affiliation{
 Sahlgrenska University Hospital, Gothenburg, Sweden
}%

\author{Armin Spreco}
\author{Toomas Timpka}
\affiliation{Department of Health, Medicine and Caring Science, Linköping
University}
\affiliation{Center for Health Services Development, Region Östergötland, Linköping, Sweden}

\author{Anna Jöud}

\affiliation{Department of Laboratory Medicine, Lund University, Sweden
}%
\affiliation{Skåne University Hospital, Department of Research and Development, Lund, Sweden}%

 \author{Torbjörn Lundh}%
\affiliation{Mathematical Sciences, Chalmers University of Technology and University of Gothenburg, Sweden}

\date{\today}

\begin{abstract}
The transmission of COVID-19 is dependent on social contacts, the rate of which have varied during the pandemic due to mandated and voluntary social distancing. Changes in transmission dynamics eventually affect hospital admissions and we have used this connection in order to model and predict regional hospital admissions in Sweden during the COVID-19 pandemic. We use an SEIR-model for each region in Sweden in which the infectivity is assumed to depend on mobility data in terms of public transport utilisation and mobile phone usage. The results show that the model can capture the timing of the first and beginning of the second wave of the pandemic. Further, we show that for two major regions of Sweden models with public transport data outperform models using mobile phone usage. The model assumes a three week delay from disease transmission to hospitalisation which makes it possible to use current mobility data to predict future admissions.
\end{abstract}

\maketitle


\section{Introduction}
Infectious diseases are disseminated through transmission of infectious agents in association with physical meetings (social contacts) between individuals. These meetings occur at home or at other locations such as workplaces or schools, which are reached using some means of transportation, e.g. by car, public transport or foot. The meetings tend to take on regular patterns and variations, and these can be used for different types of analytic purposes \cite{Holm2007,Stromgren2017}.

The COVID-19 pandemic has affected the society
in numerous ways. One striking feature is the reduction in individual mobility, which has been enforced either by strict legal lockdowns or, as in the case of Sweden, by recommendations to the general public. This reduction in mobility has had the intended effect of ``flattening the curve'' during the first, second and possibly future waves of the pandemic. 

Obtaining an understanding of the effect of mobility on the transmission of COVID-19 requires an ability to measures and quantify said changes. This has been achieved by geographically tracking cell phone usage , either directly by mobile phone operators \cite{Klein2020} or via usage of Google services \cite{google2020} that are readily available for all regions. In addition to this, mobility has also been measured by considering the utilisation of public transport \cite{jelenius2020}.

This type of information has been used in a number of studies in order to model and understand the pandemic. Linka et al. used mobility data to obtain a correlation between the reproduction number and public health interventions \cite{Linka2020}, while Zhou et al. investigated the delay of outbreaks caused by mobility restrictions \cite{Zhou2020}. Another application of mobility data is to make model-informed choices between different reopening strategies \cite{Liu2020}. 

Given the time delay between initial infection and potential hospital treatment, mobility data, such as records of daily commuters, also offer an opportunity to make predictions about the coming number of cases \cite{Picchiotti2020}. Such models can be useful for hospital administration since it allows for planning and a higher degree of preparedness for coming surges in the need of hospital beds. The aim of this study was to investigate whether variations in data reflecting weekly commuting rates were associated with later COVID-19 hospitalisation rates and also compare the ability of different data sources to achieve this aim. The underlying assumption is that decreased levels of local commuting reflect a corresponding decrease in COVID-19 transmission.


\section{Methods}
A retrospective design was used for data collection and analysis. We developed an SEIR-model of disease transmission which outputs the expected number of hospital admissions. Here we describe the hospital admission and mobility data, the epidemiological model that we have used as well as the method for fitting the model to data. The code for the model and the data used is available at: \url{http://www.math.chalmers.se/~gerlee/SEIR.html}

\subsection{Data}
Endpoint data: We consider hospital admission data from Sweden at the regional level aggregated by National Board of Health and Welfare \cite{socstyr}. The data contains the total number of newly admitted patients diagnosed with COVID-19 per week, starting with week 10. The data is reported separately for each of the 21 regions in Sweden. Missing data points were replaced by zeroes for all regions.

Syndromic data: In order to account for changes in behaviour due to governmental recommendations we have made use of mobility data from two sources: public transport data from the public transport authorities in Region Västra Götaland and Region Skåne called Västtrafik (VT) and Skånetrafiken (ST), and Google mobility reports (GMR). The VT- and ST-data describe the total number of journeys made by public transport in the region and are reported on a weekly basis. Data are given in terms of a percent change compared to travel during week 9. The GMR-data also describes the change in mobility compared to a baseline, which is the median value from the 5‑week period Jan 3 – Feb 6, 2020. Mobility is split into place categories and we have used values from the category 'transit stations'. The GMR-data is reported on a daily basis and in order to make it compatible with the model we calculate weekly averages. Figure \ref{fig:mob} in the Appendix shows the above mobility measures as a function of time.

\subsection{Epidemiological model}
To model the weekly time series of COVID-19 related hospital admissions we have used an SEIR-model with time-dependent infectivity $\beta(t)$ which is informed by mobility measures. Infectivity is assumed to vary with mobility such that the number of new social contacts for each infected individual increases with travel.

We assume that mobility measured by public transport utilisation and mobile phone usage reflects the general level of mobility in each region, which is then assumed to impact the contact rate and consequently the infectivity. Note that we do not assume that disease transmissions occurs exclusively during travel, but rather that the above mobility measures serve as a useful proxy for the rate of social contacts. 

The model is defined in terms of the following set of coupled ordinary differential equations:

\begin{equation}
    \left\{
\begin{array}{rcl}
\dfrac{dS}{dt} &= &-\dfrac{\beta(t) SI}{N}\\[8pt] 
\dfrac{dE}{dt} &= &\dfrac{\beta(t) SI}{N} - \rho E\\[8pt]  
\dfrac{dI}{dt} &= &\rho E - \gamma I \\[8pt]  
\dfrac{dR}{dt} &= &\gamma I.
\end{array}
\right.
\label{eq:SEIR}
\end{equation}

Here $\rho$ is the rate at which people leave the exposed compartment, $\gamma$ is the rate of recovery and $N$ is the population size of the region. In order to solve the system we also need to specify an initial condition and when in time it occurs. We assume that all individuals are susceptible except an initial number of $I_0$ of infectious individuals at $t_0$ weeks prior to the first data point in the admission data (week 10).

To connect the dynamics of the SEIR-model with hospital admissions we assume that individuals in the infectious compartment give rise to future hospital admissions. To model this we assume that the number of hospital admissions $t_a$ weeks into the future is given by a fraction $p$ of the present number of infectious individuals.

\subsection{Model parametrisation and fitting}
The parameters of the SEIR-model were taken from previously published studies and we have used $\rho=1.37$ week$^{-1}$ (corresponding to a latency period of on average 5.1 days) and recovery rate $\gamma=1.4$ week$^{-1}$ (corresponding to a infectious period of on average 5 days) \cite{FHM}.

Since testing was limited during the early stages of the pandemic in Sweden it is difficult to estimate the initial condition for our model. For simplicity we assume a single infected individual in a population of susceptibles appearing $t_0=4$ weeks prior to the first data point. Adjusting the initial condition for each region could possibly yield more accurate prediction, but here we have chosen a robust initial condition which gives sensible predictions for all regions.

The scaling that relates the number of infected to hospital admissions was set to $p=0.023$ in accordance with a previous study \cite{FHM}. The time lag from infection to hospital admission was set to $t_a = 3$ weeks. This value is related to the time from infection to hospital admission, which has been reported to be 17 days (5 days latency \cite{FHM} plus 12 days from symptom onset to admission \cite{zhou2020clinical}). However, 
it should not be interpreted as a parameter describing the fate of an individual patient, but should rather be interpreted as the time it takes for changes in disease transmission to propagate (sometimes via secondary cases) to hospital admissions. A previous study using mobility data has shown a time delay in admissions due to mobility restrictions in the range of 9-25 days \cite{vinceti2020}, which covers our assumed value of 21 days.

Given the uncertainty in many of the above parameter values we have carried out a sensitivity analysis by varying one parameter at a time within a reasonable range. The results of this analysis is presented in the Appendix.

The infectivity $\beta(t)$ is informed by the mobility data in the following way: For Västra Götaland and Skåne we use the public transport data and assume a linear relationship
\begin{align*}
    \beta(t) = a+bV(t)
\end{align*}
where $a,b$ are parameters that are fitted to the admission data (see below for details) and $V(t)$ is the change in travel during week $t$. For all other regions we use the GMR-data in a similar way and assume that

\begin{align*}
    \beta(t) = a+bG_i(t)
\end{align*}
where $G_i(t)$ is the GMR-data (place category 'transit stations') for region $i$, and $a,b$ are parameters that are estimated. 




In order to account for the fact that not only mobility changed at the onset of the pandemic, but also other circumstances such as physical distancing and increased hand hygiene, we adjust the baseline values for $V(t)$ and $G_i(t)$ from 0 to 0.2. 

The infectivity parameters $a,b$ are estimated by minimising the mean squared error (RMSE)

\begin{align*}
    E(\theta) = \sqrt{\frac{1}{n}\sum_{i=0}^n \left(pI(t_i+t_a,\theta)-A(t_i) \right)^2}
\end{align*}
with respect to $\theta=(a,b)$. Here $pI(t_i+t_a,\theta)$ is the predicted number of hospital admissions and $A(t_i)$ is the actual number of admissions and the sum runs over all time points $t_i$. To find the minimum RMSE we use the grid search method  with 80 linearly spaced values in the range 1-12 for both $a$ and $b$ \cite{chang2020}. For each region $i$ we thus obtain a set $\hat{\theta}_i=(\hat{a}_i,\hat{b}_i)$ of estimated parameters. When comparing the model error between different regions we normalise the RMSE by dividing with the maximum number of weekly admissions for each region.

In order to quantify the uncertainty in our parameter estimates we select all parameter sets $(a,b)$ that achieve an RMSE of within 20\% of $E(\hat{\theta})$. We solve the SEIR-model for all those parameter combinations and remove the lower and upper 5th percentile to obtain a 95\% credible interval. This procedure corresponds to sampling from the posterior in an Approximate Bayesian Computation framework with $E(\theta)$ as our summary statistic \cite{alahmadi2020}.

For Region Västra Götaland we fit the mobility-driven SEIR-model (\ref{eq:SEIR}) using increasing amounts of reported hospital admissions. We start by including data up until week 20 and test the models predictive ability in terms of the mean average predictive error (MAPE) on the coming three weeks. This procedure is repeated for increasing amounts of training data. To illustrate the robustness of the model we also plot how the estimated model parameters $\hat{a}$ and $\hat{b}$ change as we include more weekly data.
\section{Results}
\subsection{Predicting hospital admissions using public transport utilisation}
\begin{figure}
\includegraphics[width=\textwidth]{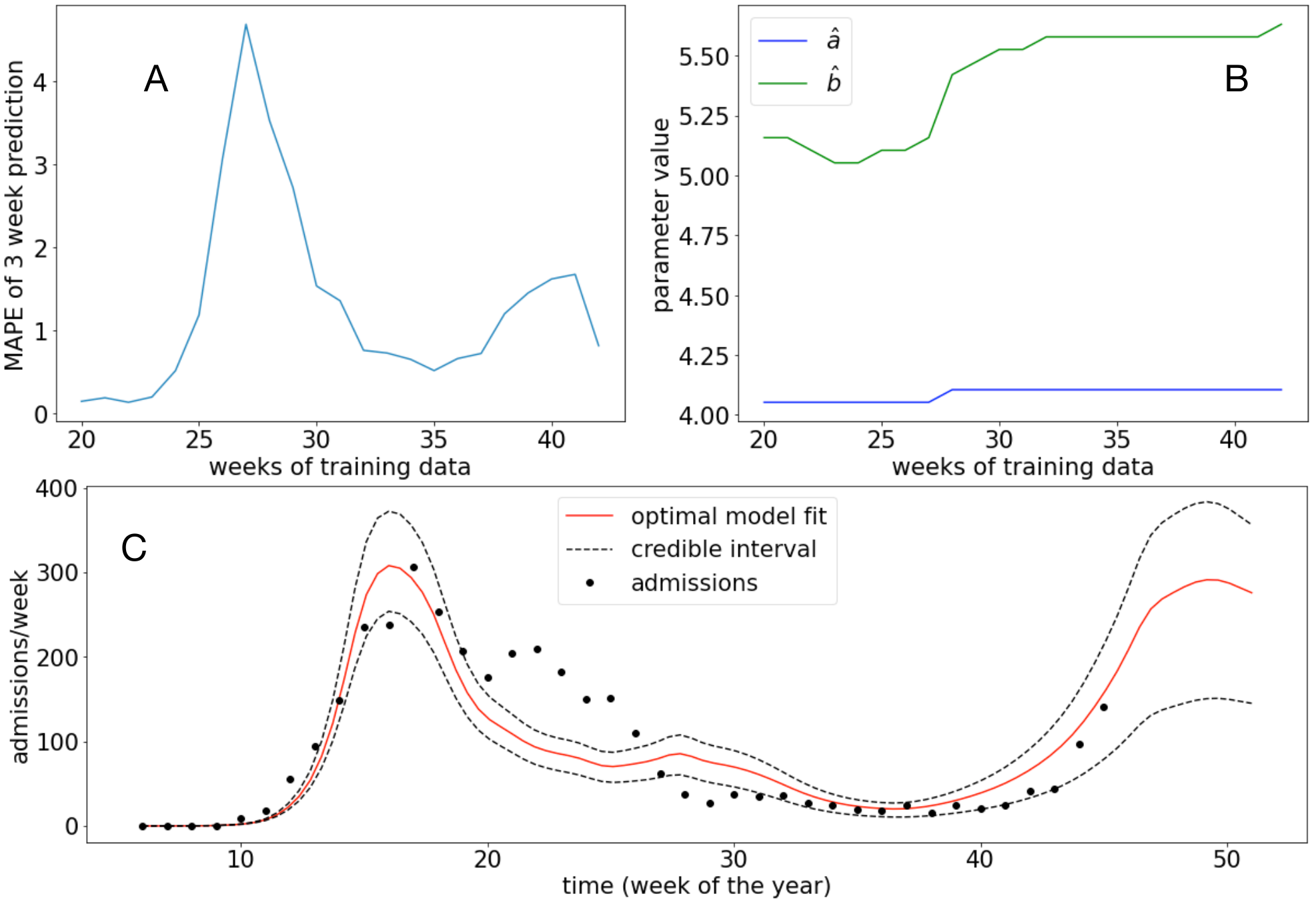}
\caption{\label{fig:VGR} Model fit to admission data from Region Västra Götaland. \textbf{A} The model error in terms of the MAPE on 3 week predictions as a function of the number of weeks of data used in the fitting. \textbf{B} The estimated model parameters $(\hat{a},\hat{b})$ as a function of the number of weeks of data used in the fitting. \textbf{C} The optimal fit when all data points are used (until week 45). The dashed lines show the 95\% credible interval for the model fit (see Methods). }
\end{figure}

For Region Västra Götaland the resulting model error in terms of MAPE can be seen in figure \ref{fig:VGR}A. By successively increasing the training data, we see in fig.\ \ref{fig:VGR}B  that the model remains largely unchanged beyond week 30, which timewise corresponds to the end of the first wave of the pandemic.

When using all available data we find that $\hat{a}=4.16$ and $\hat{b}=5.74$ (fig.\ \ref{fig:VGR}C), and we note that the model captures the dynamics of admissions during both the first and beginning of the second wave, although the rate of decline during the first wave is overestimated. 

\subsection{Using Google mobility data to predict hospital admissions}

For all other regions we make use of Google mobility data (see Methods for details). Figure \ref{fig:Sthlm} shows model fits for Östergötland and Stockholm (see fig.\ \ref{fig:alla} in the Appendix for model fits to admissions in all Swedish regions and table \ref{tab:params} for normalised RMSE and estimated parameters). Again, we note that the model correctly describes the timing of the first and second wave. Visual inspection of the model fits for all regions suggest that the model performs better for regions with a larger population.

\begin{figure}[!htb]
\includegraphics[width=13cm]{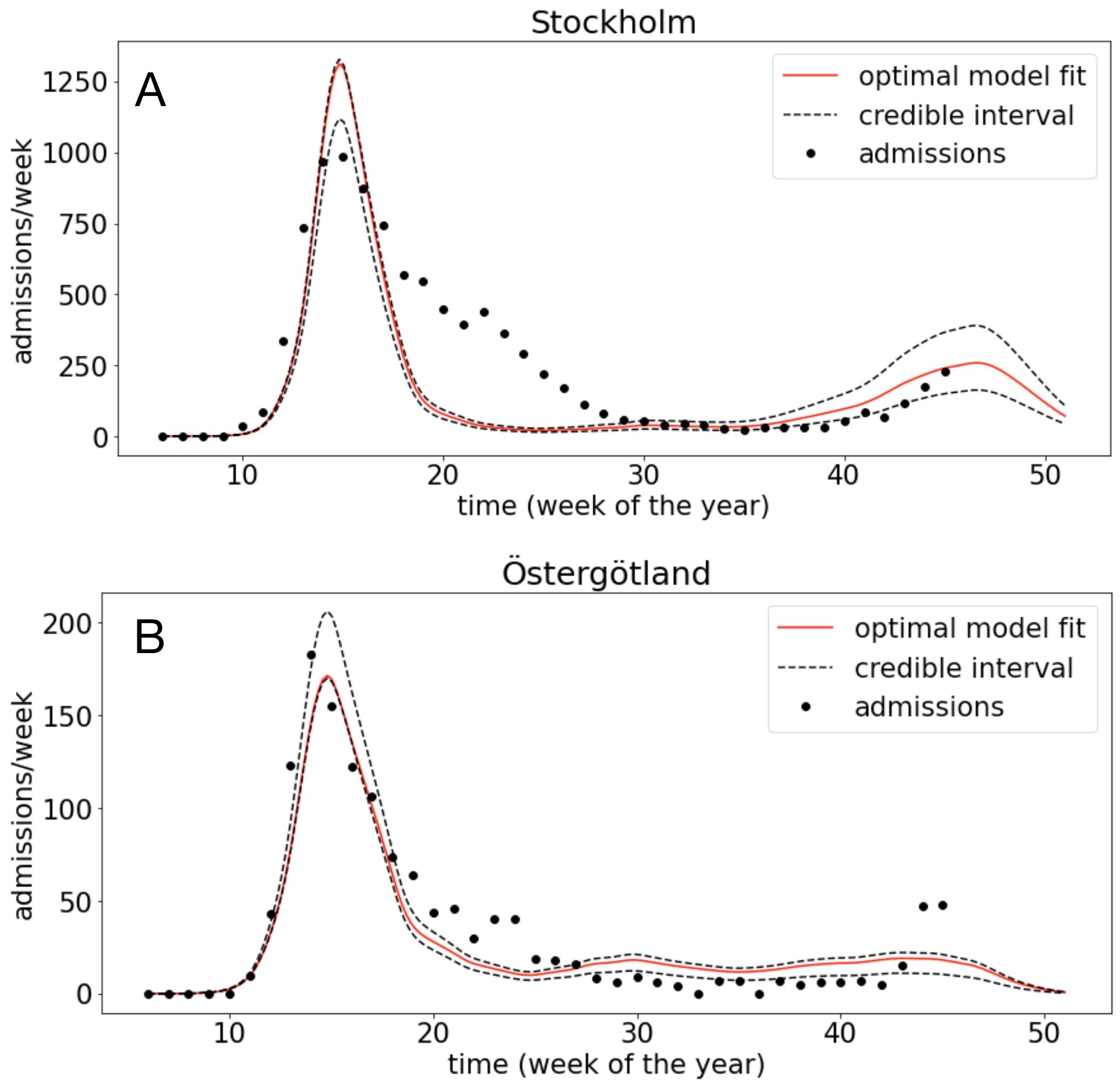}
\caption{\label{fig:Sthlm} Optimal model fit for \textbf{A} Stockholm ($\hat{a}=4.68$ and $\hat{b}=6.37$) and \textbf{B} Östergötland ($\hat{a}=3.79$ and $\hat{b}=5.89$). In both panels the dashed lines show 95 \% credible intervals for the model fit.}
\end{figure}

\subsection{Public transport data improves model fit compared to google mobility reports}

\begin{figure}[!htb]
\includegraphics[width=\textwidth]{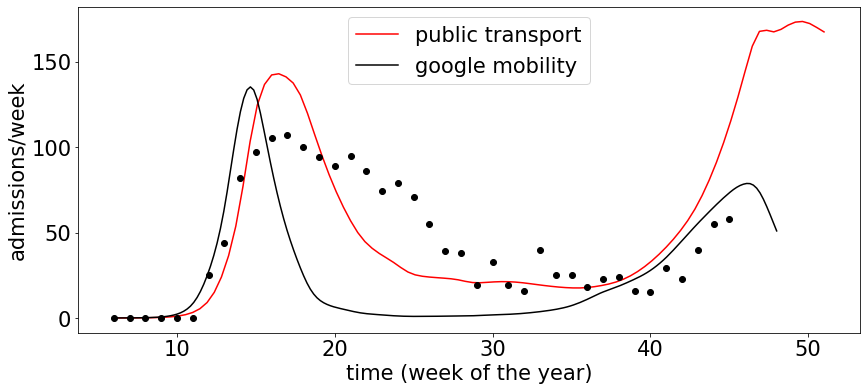}
\caption{\label{fig:Skåne} Optimal model fit for Skåne Region using mobility data from public transport (red line) and Google Mobility Report (black line).}
\end{figure}

For Skåne Region we have both public transport data and GMR-data,  which was used on all other regions. Figure \ref{fig:Skåne} shows the best model fits using the mobility data from the public transport agency Skånetrafiken compared to GMR-data. We note that although the model using GMR fits the data during the second wave better the overall fit is considerably improved by using data from public transport. In terms of the RMSE we observe that the model error is 23 admissions/week for the public transport model compared to 40 admissions/week for the model that uses GMR. A similar trend is seen for Region Västra Götaland where public transport data yields an RMSE of 40 admissions/week whereas GMR-data gives an error of 90 admissions/week.


\section{Discussion}
We set out to investigate whether variations in data reflecting weekly commuting rates were associated with later COVID-19 hospitalisation rates. It was found that COVID-19 hospital admission can be modelled using time-dependent mobility data and that a SEIR model can be fitted using two free parameters to regional data from Sweden.

Our approach is similar to a recent study by Chang et al. \cite{chang2020} who used spatially resolved mobility data in order to model disease transmission in metropolitan areas in the US. They compared their model output to COVID-19 incidence, whereas we have focused on hospital admissions. The reasons for this are twofold: firstly, the data on incidence in Sweden is unreliable due to limited testing and secondly hospital admissions is a more interesting metric for healthcare providers. Given the time lag between current mobility which drives infections and future hospital admissions, the model provides a tool for predicting the demands on hospital beds up to three weeks in advance. Although the SEIR-model describes the transmission of the disease, the model details were not in focus in the present study. Uncertainty in model parameters such as the initial condition and the fraction of individuals that become hospitalised implies that the model dynamics in terms of the number of susceptible, infectious and recovered individuals are unreliable. It is also worth pointing out that the connection between infection and hospitalisation is not assumed to be direct. It may well be that an individual who contributes to the measured mobility transmits the virus in several steps to an individual with an increased risk of severe illness who is subsequently hospitalised. 

Despite these simplifications, the model was able to capture the general shape and timing of both the first and the beginning of the second wave for most regions. There appears to be a link between the population size of the region and the goodness of fit. The model fits the admission data for larger regions better, and a possible explanation for this is the large degree of randomness seen in the smaller regions. A recurring feature seen across most regions is the inability of the model to accurately describe the width of the first peak. The model tends to underestimate the actual width, and this is likely due to the lack of detail in the model or inaccuracy in the assumed parameter values. Moreover, it is noteworthy that the model captured the timing of the onset of the second peak without accounting for seasonal or temperature-driven infectivity. Instead, the results suggest that mobility in itself, which might contain seasonal variation is sufficient to capture the dynamics of hospital admissions. 

The present research has some limitations that should be taken in consideration when interpreting the results. When trained on admission data until week 20 for Region Västra Götaland the model initially performs well in terms of the model error (MAPE) on the 3 consecutive weeks (Fig. \ref{fig:VGR}A). This is followed by an increase in the MAPE during week 25-30, which is due to an increase in admissions that the model is unable to capture, and then a subsequent decline. In terms of the model robustness, we observe that the parameters remain largely unchanged after week 30, suggesting that data from the first wave was sufficient to fit the model (Fig. \ref{fig:VGR}B). 

In the model we have disregarded any kind of age-structure, migration of cases between regions and assumed a highly simplified connection between infection and hospital admission. In addition, we have assumed that the disease was introduced in an identical way in all regions. These choices were made in order to formulate a simple and general model, which could be applied directly to all regions. We have shown for Region Skåne that public transport data provides a better fit between model and admission data, and further tailoring the model to each region will most likely improve model fit even further. In terms of numerical methods, we have also made a couple of simplifications. We carried out the parameter estimation one region at a time. Here it would be beneficial to consider a hierarchical mixed-effects model that considers all regions simultaneously \cite{lindstrom1990}. We have performed a sensitivity analysis with respect to the initial condition, the parameters that relate the size of the infectious compartment to hospital admissions s and the initial mobility (see Fig. \ref{fig:sens}). The results show that the model fit can be somewhat improved by making slight adjustment to the baseline parameter values. However, given the uncertainty in these parameter values we do not find it motivated to adjust our baseline values. 

This study should be seen as a first attempt to model regional-level hospital admissions in Europe using mobility data. The assumed delay of three weeks between infection and admission implies that the model can, with current mobility data, make predictions three weeks into the future. The results encourage continued research on use of mobility data in health service capacity planning during the COVID-19 pandemic.

\section{Acknowledgements}
We would like to thank Jonas Hägglund at Västtrafik for providing the public transport data. PG and TL would like to acknowledge seed funding from Chalmers University of Technology Areas of Advance ``Information and Communication Technology'' and ``Health Engineering''.


%

\clearpage
\section{Appendix}

\subsection{Sensitivity analysis}
\begin{figure}[!htb]
\includegraphics[width=15cm]{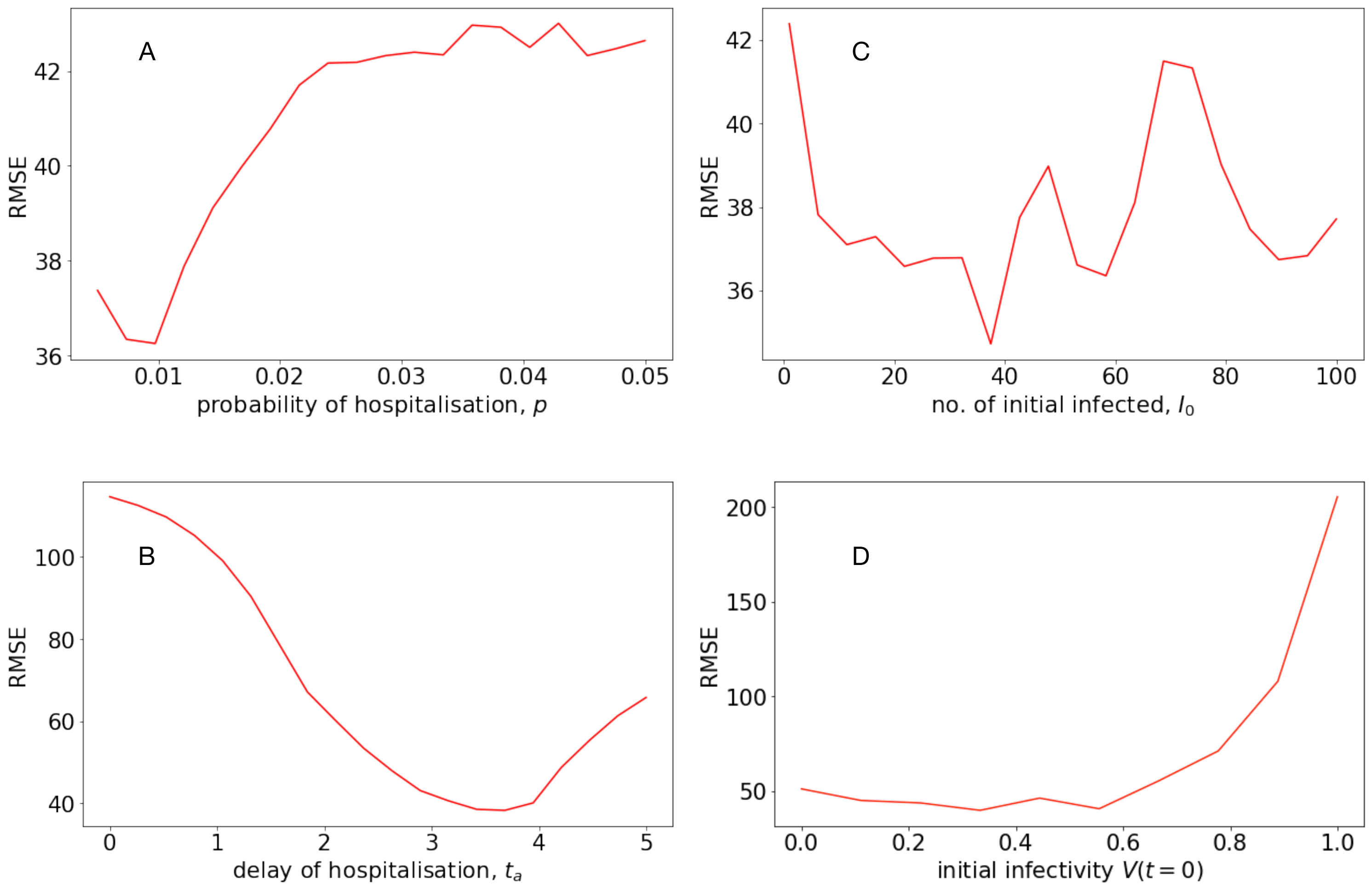}
\caption{\label{fig:sens}Sensitivity analysis of model parameters for Region Västra Götaland. The defauly values are $p=0.023$, $t_a=3$, $I_0=1$ and $V(t=0)=0.2$.}
\end{figure}

Figure \ref{fig:sens} shows how the model error (RMSE) of the best fit for Region Västra Götaland changes when the parameters $p$, $t_a$,$I_0$ and $V(t=0)$ are varied. We note that it is possible to achieve a slightly better model fit when the probability of hospitalisation is lowered to $p=0.1$, but the improvement in model fit is minor. For the delay we see that our value of $t_a=3$ weeks lies close to a local minimum, but little would be gained (in terms of RMSE) by increasing the delay. The number of infected individuals at $t=0$ has a more complicated impact on the error. A smaller RMSE could be achieved by increasing $I_0$ from its default value of 1, but the improvement is again minor. Lastly, the initial infectivity has a minor impact on the model error as long as it remains below 0.6.

\clearpage
\subsection{Fitting the model to 20 Swedish regions}
Here we present model fits for all Swedish regions expect Gotland for which no data was available from the National Board of Health and Welfare.

\begin{figure}[!htb]
\includegraphics[width=\textwidth]{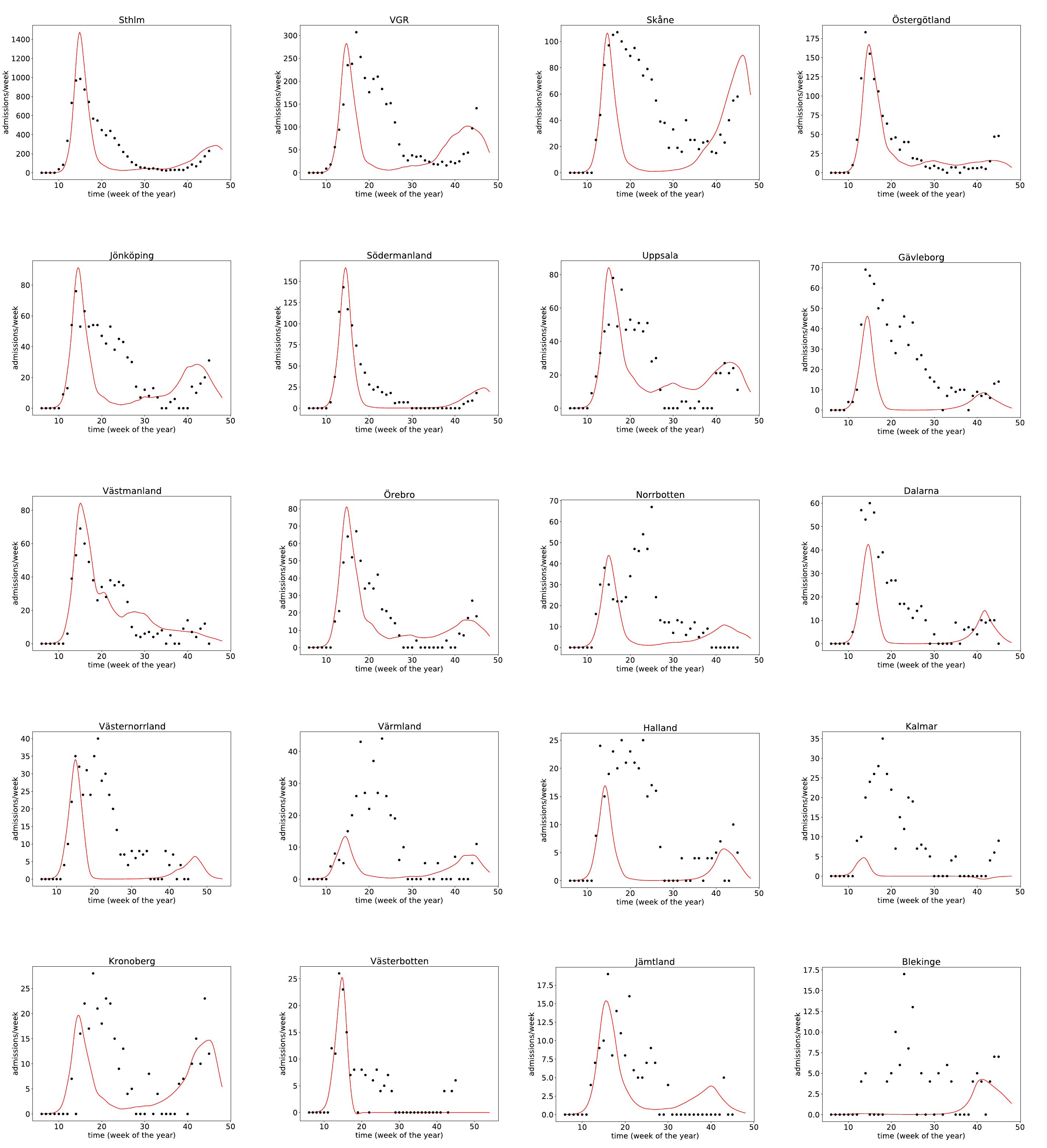}
\caption{\label{fig:alla} Optimal model fit for all Swedish regions except Gotland. Estimated parameter values can be found in table \ref{tab:params}.}
\end{figure}

\begin{table}[]
    \centering
    \begin{tabular}{|l|r|c|c|r|}
\hline
         Region &  Population, $N$ &         $\hat{a}$ &          $\hat{b}$ &         normalised RMSE \\
\toprule
          Sthlm &     2389923 & 4.76 &  9.22 & 0.18 \\
            VGR &     1725881 & 3.85 &  9.49 & 0.30 \\
          Skåne &     1387650 & 3.62 &  8.24 & 0.39 \\
   Östergötland &      467095 & 3.73 &  8.38 & 0.09 \\
      Jönköping &      364750 & 3.28 & 10.05 & 0.28 \\
   Södermanland &      299101 & 3.73 &  9.22 & 0.11 \\
        Uppsala &      387628 & 3.51 &  6.43 & 0.23 \\
      Gävleborg &      287660 & 3.16 & 10.75 & 0.30 \\
    Västmanland &      277074 & 3.51 &  6.85 & 0.12 \\
         Örebro &      305726 & 3.39 &  8.38 & 0.18 \\
     Norrbotten &      249768 & 3.05 &  7.13 & 0.31 \\
        Dalarna &      287806 & 2.94 & 10.61 & 0.20 \\
 Västernorrland &      244855 & 2.94 &  9.63 & 0.35 \\
       Värmland &      282840 & 2.71 &  9.08 & 0.34 \\
        Halland &      336132 & 2.59 & 10.19 & 0.42 \\
         Kalmar &      245992 & 2.37 & 11.16 & 0.34 \\
      Kronoberg &      202163 & 2.94 &  6.85 & 0.31 \\
   Västerbotten &      273061 & 2.94 &  7.96 & 0.13 \\
       Jämtland &      130972 & 2.82 &  5.18 & 0.21 \\
       Blekinge &      159349 & 1.68 &  1.28 & 0.27 \\
\hline
\end{tabular}

    \caption{Population size, estimated parameters and model error (normalised RMSE) for all considered regions.}
    \label{tab:params}
\end{table}

\clearpage
\subsection{Mobility data for Västra Götaland}

\begin{figure}[!htb]
\includegraphics[width=9cm]{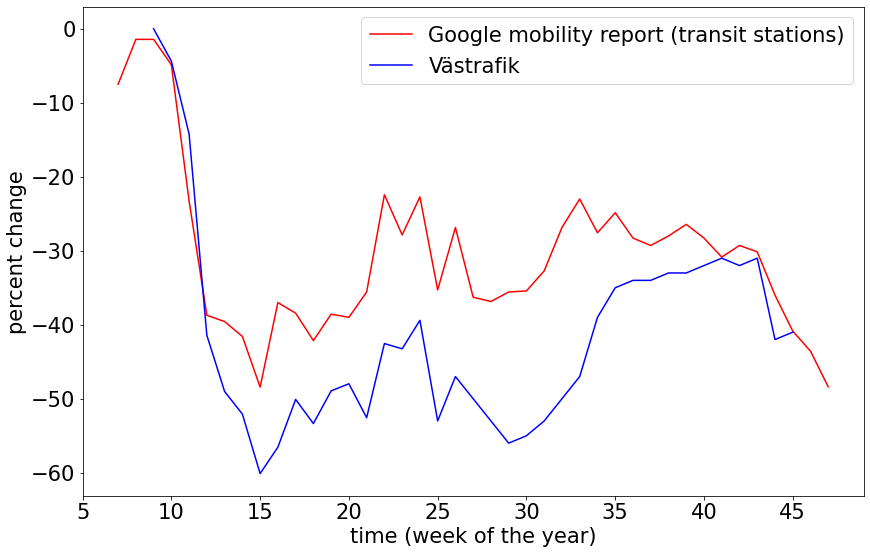}
\caption{\label{fig:mob}Mobility data for Region Västra Götaland in terms of public transport usage (blue) and Google mobility report (red). See methods for details.}
\end{figure}

\end{document}